\documentclass[preprint,prb]{revtex4}
\usepackage{amsmath}
\usepackage{graphicx}
\usepackage{dcolumn}
\usepackage{bm}

\begin{document}

\begin{titlepage}

\title
{Spin-flip effects on the Andreev bound states and supercurrents
in a superconductor/quantum-dot/superconductor system}

\author{Hui Pan$^{1}$, Tsung-Han Lin$^{1}$, and Dapeng Yu$^{1}$}
\affiliation{$^{1}$State Key Laboratory for Mesoscopic Physics
and Department of Physics, Peking University, Beijing 100871,
China}

\begin{abstract}
We investigate the spin-flip effects on the Andreev bound states
and the supercurrent in a
superconductor/quantum-dot/superconductor system theoretically.
The spin-flip scattering in the quantum dot can reverse the
supercurrent flowing through the system, and one $\pi$-junction
transition occurs. By controlling the energy level of quantum
dot, the supercurrent is reversed back and another $\pi$-junction
transition appears. The different influences of the spin-flip
scattering and the intradot energy level on the supercurrent are
interpreted in the picture of Andreev bound states.

\noindent {PACS number(s)}: 74.50.+r, 73.23.-b, 73.63.Kv
\end{abstract}
\maketitle
\end{titlepage}

\section{Introduction}
The superconductor coupled mesoscopic hybrid systems have
attracted much attention in recent years, not only because of
fundamental interest, but also of potential applications for
future nanoelectronics.\cite{Ralph,Knorr,Buitelaar,Takayanagi} In
ballistic superconductor/normal-metal/superconductor (S/N/S)
junction, Andreev bound states can be formed.\cite{Bagwell} Each
Andreev bound state carries a supercurrent in positive or negative
direction at a given phase difference $\phi$ between the two
superconductors. Therefore, the net supercurrent between two
superconductors depends not only on the phase difference $\phi$,
but also on the occupation of the Andreev bound states. When two
superconductors are weakly linked, the current-phase relation is
$I=I_{c}\sin(\phi)$. On some occasions, the sign of $I_{c}$ may be
reversed,\cite{Baselmans} which is referred to as the
$\pi$-junction transition, because the minus sign can be absorbed
in to the phase factor as $\sin(\phi+\pi)$.

The spin-orbit interaction plays an important role in the quantum
dot (QD), because it can change the spin orientation of an
electron. The spin-flip mechanisms in the GaAs-based QD have been
studied.\cite{Khaetskii} The spin-flip effects on trasport
properties of a quantum dot in the normal-metal and
superconductor hybrid system has been studied.\cite{Xing} Various
resonant peaks appear for the different spin-flip strengths in
the QD.\cite{Xing} The spin-dependent Andreev reflection
tunneling through a QD with the spin-flip scattering has also
been studied.\cite{Chen} It is found that competition between the
intradot spin-flip scattering and the tunneling coupling to the
leads dominates the resonant behaviors of the Andreev
reflection.\cite{Chen}

It is natural to ask if the intradot spin-flip scattering could
induce some novel phenomena in the supercurrent such as the
$\pi$-junction transition. Motivated by this, we investigate the
spin-flip scattering effects on the supercurrent and Andreev
bound states in the superconductor/quantum-dot/superconductor (
S/QD/S) system in this paper. By using the standard nonequilibrium
Green's function (NGF)
techniques,\cite{Yeyati1,Yeyati2,Sun1,Sun2} we have analyzed
quantum transport properties of the S/QD/S system. The
configuration of the Andreev bound states depend heavily on the
spin-flip strength. Since the supercurrent is carried by these
states, the spin-flip scattering has a great influence on the
amplitude and sign of the supercurrents. The dependence of the
supercurrent and Andreev bound states on the gate voltage is also
studied.

\section{Physical model and formula}
The S/QD/S system under consideration is described by the
following Hamiltonian:
\begin{equation}
H=\sum_{\alpha=L,R}H_{\alpha}+H_{dot}+H_{T},
\end{equation}
with
\begin{equation}
H_{\alpha}=\sum_{k,\sigma}\epsilon_{\alpha,k}a_{\alpha,k\sigma}^{\dag}a_{\alpha,k\sigma}
+\sum_{k}[\Delta e^{-i\phi_{\alpha}}
a_{\alpha,k\uparrow}^{\dag}a_{\alpha,-k\downarrow}^{\dag}+H.c.],
\end{equation}
\begin{equation*}
H_{dot}=\sum_{\sigma}\epsilon_{0}d_{\sigma}^{\dag}d_{\sigma}
 +r(d_{\uparrow}^{\dag}d_{\downarrow}+d_{\downarrow}^{\dag}d_{\uparrow}),
\end{equation*}
\begin{equation*}
H_{T}=\sum_{\alpha,k\sigma}
[t_{\alpha}a_{\alpha,k\sigma}^{\dag}d_{\sigma}+H.c.],
\end{equation*}
where $H_{\alpha}$ ($\alpha=L/R$) is the standard BCS Hamiltonian
for the left/right superconducting leads with phase
$\phi_{L}/\phi_{R}$ and the energy gap $\Delta$. $H_{dot}$ models
the quantum dot with a single spin-degenerate level
$\epsilon_{0}$, which can be controlled by the gate voltage. The
spin-flip term in the $H_{dot}$ comes from the spin-orbit
interaction in the quantum dot. $H_{T}$ denotes the tunneling
part of the Hamiltonian, and $t_{L,R}$ are the hopping matrix.
The supercurrent can be calculated from standard NGF techniques.

The $4\times 4$ Nambu representation is used to include the
physics of AR and the spin-flip process in a unified formalism.
The retarded Green's function if defined as
$G^{r}_{\alpha,\beta}(t,t')= \mp i\theta(\pm t\mp
t')\langle\{\Psi_{\alpha}(t),\Psi_{\beta}^{\dag}(t')\}\rangle$
with the operator
$\Psi_{\alpha}=(\psi_{\alpha\uparrow}^{\dag},\psi_{\alpha\downarrow}
 ,\psi_{\alpha\downarrow}^{\dag},\psi_{\alpha\uparrow})^{\dag}$.
Let $g^{r}$ and $G^{r}$ denote the Fourier-transformed retarded
Green's function of the QD without and with the coupling to the
leads. In the Nambu representation, they can be written as
\begin{equation}
(g^{r}(\epsilon))^{-1}=\left(\begin{array}{cccc}
 \epsilon-\epsilon_{0\uparrow}+i0^{+} & 0 & -r & 0\\
 0 & \epsilon+\epsilon_{0\downarrow}+i0^{+} & 0 & r\\
 -r & 0 & \epsilon-\epsilon_{0\downarrow}+i0^{+} & 0\\
 0 & r & 0 & \epsilon+\epsilon_{0\uparrow}+i0^{+}
\end{array}\right).
\end{equation}
The retarded self-energy under the wide-bandwidth approximation
can be derived as\cite{Sun1,Sun2}
\begin{equation}
\Sigma^{r}_{L/R}(\epsilon)=
-i\Gamma_{L/R}\rho(\epsilon)\left(\begin{array}{cccc}
 1 & -\frac{\Delta}{\epsilon}e^{-i\phi_{L/R}} & 0 & 0\\
 -\frac{\Delta}{\epsilon}e^{i\phi_{L/R}} & 1  & 0 & 0\\
 0 & 0 & 1 & \frac{\Delta}{\epsilon}e^{-i\phi_{L/R}}\\
 0 & 0 & \frac{\Delta}{\epsilon}e^{i\phi_{L/R}} & 1
\end{array}\right),
\end{equation}
where $\Gamma_{L/R}$ is the appropriate linewidth functions
describing the coupling between the dot and the respective
superconducting leads. Under the wide-bandwidth approximation,
the linewidth functions are independent on the energy variable.
Furthermore, we set $\phi_{L}=\phi/2$ and $\phi_{R}=-\phi/2$,
$\Gamma_{L}=\Gamma_{R}=\Gamma$ with small values for the
symmetric and weak-coupling case. The factor $\rho(\epsilon)$ is
defined as
\begin{equation}
\rho(\epsilon)= \left\{ \begin{array}{cc}
 \frac{|\epsilon|}{(\epsilon^{2}-\Delta^{2})} & |\epsilon|>\Delta\\
 \frac{|\epsilon|}{i(\Delta^{2}-\epsilon^{2})} & |\epsilon|<\Delta
\end{array} \right..
\end{equation}
By using the Dyson equation, the retarded Green function of the
system can be obtained as
\begin{equation}
G^{r}(\epsilon)=[g^{r^{-1}}(\epsilon)-\Sigma^{r}(\epsilon)]^{-1},
\end{equation}
where
$\mathbf{\Sigma}^{r}=\mathbf{\Sigma}^{r}_{L}+\mathbf{\Sigma}^{r}_{R}$.
The Josephson current is expressed as
\begin{equation}
I_{L/R}=I_{L/R,\uparrow}+I_{L/R,\downarrow}
 =\frac{2e}{\hbar}\int\frac{d\epsilon}{2\pi}Tr\{\hat{\sigma}_{z}Re[G\Sigma_{L/R}]^{<}\},
\end{equation}
where $[AB]^{<}\equiv A^{<}B^{a}+A^{r}B^{<}$ and
$\hat{\sigma}_{z}$ is a $4\times 4$ matrix with Pauli matrix
$\sigma_{z}$ as its diagonal components. In the steady transport,
the current is
\begin{equation}
I=\frac{1}{2}(I_{L}-I_{R})
 =\frac{e}{\hbar}\int\frac{d\epsilon}{2\pi}j(\epsilon)
 =\frac{e}{\hbar}\int\frac{d\epsilon}{2\pi}Tr\{\hat{\sigma}_{z}
 Re[G(\Sigma_{L}-\Sigma_{R})]^{<}\}.
\end{equation}
Applying the fluctuation-dissipation theorem, one has
\begin{equation}
G^{<}(\epsilon)=f(\epsilon)(G^{a}(\epsilon)-G^{r}(\epsilon)),
\quad
\Sigma^{<}_{L/R}=f(\epsilon)(\Sigma^{a}_{L/R}-\Sigma^{r}_{L/R}),
\end{equation}
where $f(\epsilon)=1/(e^{\beta\epsilon}+1)$ is the Fermi
distribution function. Consequently, the Josephson current is
expressed as
\begin{equation}
I=\frac{e}{\hbar}\int\frac{d\epsilon}{2\pi}f(\epsilon)j(\epsilon),
\end{equation}
in which the current density $j(\epsilon)$ is defined as
\begin{equation}
j(\epsilon)=Tr\{\hat{\sigma}_{z}Re[G^{a}(\Sigma^{a}_{L}-\Sigma^{a}_{R})
 -G^{r}(\Sigma^{r}_{L}-\Sigma^{r}_{R})]\},
\end{equation}
The analysis of the current carrying spectrum $j(\epsilon)$
provides the information of the supercurrent carried by each of
the Andreev bound state. The Josephson current can be divided
into two parts, contributed by the continuous spectrum for
$|\omega|>\Delta$ and discrete spectrum for $|\omega|<\Delta$:
\begin{equation}
I=I_{c}+I_{d},
\end{equation}
\begin{equation*}
I_{c}=\frac{e}{\hbar}(\int_{-\infty}^{-\Delta}+\int_{\Delta}^{\infty})
 \frac{d\epsilon}{2\pi}f(\epsilon)j(\epsilon),
\end{equation*}
\begin{equation*}
I_{d}=\frac{e}{\hbar}\int_{-\Delta}^{\Delta}
 \frac{d\epsilon}{2\pi}f(\epsilon)j(\epsilon).
\end{equation*}
The averaged LDOS is given by
\begin{equation}
D(\epsilon)=-\frac{1}{\pi}Tr\{Im[G^{r}(\epsilon)]\},
\end{equation}
We perform the calculations at zero temperature in units of
$\hbar=e=1$. The energy gap of the superconductor is fixed as
$\Delta=1$. All the energy quantities in the calculations are
scaled by $\Delta$. The linewidth is $\Gamma=0.1\Delta$ for the
symmetric and weak-coupling case.

\section{Results and discussion}
In the following, the numerical results of the supercurrents and
the Andreev quasibound states are discussed in detail. The
supercurrent originates from Andreev reflection at the interface
between the superconducting leads and the central region. Fig.
1(a) and (b) present the supercurrent $I$ (includes both $I_{c}$
and $I_{d}$) versus the phase difference $\phi$ with different
spin-flip scattering strength $r$. First, we investigate the case
without the spin-flip scattering $r=0$. The current $I_{d}$ from
the discrete spectrum vs $\phi$ is similar to a $\sin(\phi)$-like
curve. However, the current $I_{c}$ from the continuous spectrum
vs $\phi$ is similar to a $\sin(\phi+\pi)$-like curve, because the
current $I_{c}$ is a $\pi$-junction Josephson
relation\cite{Zabel,Harlingen}. Furthermore, the current $I_{d}$
is much larger than the current $I_{c}$, which means that the
total supercurrent $I$ is mainly contributed by $I_{d}$ and also
shows a sine-like dependence on the phase $\phi$. Now, we
investigate the case with the spin-flip scattering $r=0.2$. It is
interesting to point out that $I$ vs $\phi$ is not a
$\sin(\phi)$-like but a $\sin(\phi+\pi)$-like curve. There is a
$\pi$-junction transition for the supercurrent-phase relation.
The reason is related to the spin-flip scattering which greatly
suppresses the current $I_{d}$. Thus the total supercurrent is
mainly contributed by $I_{c}$ but not by $I_{d}$ as that without
the spin-flip scattering. Due to the quite different dependence of
$I_{d}$ and $I_{d}$ on $\phi$, the supercurrent $I$ shows a
$\pi$-junction transition under the influence of the spin-flip
scattering effects.

To fully understand the $\pi$-junction transition, we plot the
$j(\epsilon)$ for the cases with and without the spin-flip
scattering in Fig. 1(c) and (d). For the case with $r=0$, the
original level $\epsilon_{0}=0$ is split into two Andreev
quasibound states. When the energy $\epsilon$ of an incoming
electron lines up with the Andreev bound states, a resonance
occurs, leading to a very large supercurrent. As seen from Fig.
1(c), $j(\epsilon)$ has two $\delta$-function-type discrete
spectra within the superconducting gap, corresponding to the two
Andreev bound states. They carry supercurrents with opposite
signs: positive for $A_{1}$ and negative for $A_{-1}$.
$j(\epsilon)$ also has a continuous spectrum outside the
superconducting gap: negative for $C_{-1}$ and positive for
$C_{1}$. At zero temperature, only the spectrum of $\epsilon<0$
relates to the current. Since the contribution from the discrete
spectrum $A_{1}$ is much larger than that from the continuous one
$C_{-1}$, the current $I$ is mostly contributed by $I_{d}$. For
the case with $r=0.2$, the original level $\epsilon_{0}=0$ is
spit into two ones as $\epsilon_{01}$ and $\epsilon_{02}$ due to
the spin-flip perturbation, which results in four Andreev bound
states. The original Andreev bound state $A_{1}$ is split into
two ones as $A_{1}$ and $A_{2}$, and similarly $A_{-1}$ is split
into $A_{-1}$ and $A_{-2}$. $j(\epsilon)$ has four
$\delta$-function-type discrete spectra within the
superconducting gap. The Andreev bound states depend strongly on
the configuration of the QD levels
($\epsilon_{01}$,$\epsilon_{02}$), but weakly on the phase
difference $\phi$, and the electron levels $\epsilon_{01}$ and
$\epsilon_{02}$ are coupled by Andreev reflection tunneling.
Therefore, Andreev bound states can be viewed as hybrids of
$\epsilon_{01}$ and $\epsilon_{02}$. With increasing $r$, $A_{1}$
and $A_{-2}$ move in the same direction, and both of them are
below the Fermi level at strong enough $r$. As a consequence,
they make little net contribution to the supercurrent, and the
relatively small negative continuous spectrum of $C_{-1}$
dominates. The contribution from the discrete spectrum is
suppressed by the spin-flip scattering. This is the origin of the
$\pi$-junction transition in the supercurrent under the influence
of the spin-flip scattering in the QD. Another important
quantity, the local density of states (LDOS) $D(\epsilon)$, is
also shown in Fig. 1(e) and (f). A series of very narrow peaks
emerge in $D(\epsilon)$, clearly indicating the formation of
Andreev quasibound states inside the QD. The peaks of the current
density $j(\epsilon)$ are located precisely at the energies of
Andreev bound states, which is a clear indication that the
current is carried by these states.

The results above are obtained by fixing the intradot level
$\epsilon_{0}$ to zero, which is just at the center of the gap
and at the Fermi level of both left and right leads. Next, we
investigate how the supercurrent is affected when $\epsilon_{0}$
is moved away from zero by the gate voltage. The supercurrent at
different $\epsilon_{0}$ are plotted in Fig. 2. As seen from Fig.
2(a), the supercurrent at $\epsilon_{0}=0.1$ has similar
$\sin(\phi+\pi)$-like phase dependence with that in Fig. 1(b).
However, when $\epsilon_{0}=0.3$, there appears a $\pi$-junction
transition, and then the supercurrent has has similar
$\sin(\phi)$-like phase dependence with that in Fig. 1(a). To
explain this transition, the corresponding current density
$j(\epsilon)$ are also shown in Fig. 2(c) and (d). With
$\epsilon_{0}\neq 0$, the Andreev quasibound states in
$j(\epsilon)$ are shifted in their positions. The two successive
states are shifted in opposite directions. The two bound states
$A_{1}$ and $A_{2}$ carrying the positive current move to
$-\Delta$, while $A_{-1}$ and $A_{-2}$ carrying the negative
current move to $\Delta$. At small $\epsilon_{0}$, the net current
from discrete spectrum $A_{1}$ and $A_{-2}$ are very small and the
continuous spectrum $C_{-1}$ mainly contributes the current. At
large enough $\epsilon_{0}$, the position of $A_{2}$ and $A_{-2}$
can even move past over the Fermi level, and both of their
positions are below the Fermi level. Now, the current are mainly
contributed by the discrete spectrum of $A_{1}$ and $A_{2}$, but
not by $C_{-1}$ anymore. This is just the reason why the
$\pi$-junction transition occurs. The corresponding $D(\epsilon)$
shown in Fig. 2(e) and (f) are symmetric about the Fermi level.
The narrow peaks with different height clearly indicate shift of
the Andreev bound states.

To clearly show the $\pi$-junction transitions mentioned above,
the dependence of $I(\phi=\pi/2)$ on $r$ and $\epsilon_{0}$ are
plotted in Fig. 3. As shown in Fig. 3(a), $I(\pi/2)$ decreases
slowly first with increasing $r$. When the spin-flip scattering
strength $r$ is comparable and close to $\epsilon_{0}$,
$I(\pi/2)$ decreases rapidly and can even change the sign from
positive to negative. Then the $pi$-phase transition occurs as
shown in Fig. 1. At larger $\epsilon_{0}$, the stronger $r$ is
needed to move the bound state $A_{-2}$ to below the Fermi level
as $A_{1}$. The negative current carried by $A_{-2}$ counteracts
with the positive current carried by $A_{1}$. Then the
$\pi$-junction transition can occur. When there is no spin-flip
scattering $r=0$, increasing $\epsilon_{0}$ can decrease
$I(\pi/2)$ as shown in Fig. 3(b). Because the intradot energy
level $\epsilon_{01}$ and $\epsilon_{02}$ are not symmetric about
the Fermi level at nonzero $\epsilon_{0}$, the Andreev reflection
are suppressed and then the current decreases. For the nonzero
$r$, $I(\pi/2)$ first increases and then decreases. At small
spin-flip strength $r$, increasing $\epsilon_{0}$ does not change
of the sign of $I(\pi/2)$. At some strong enough $r$, which
already induces the $\pi$-junction transition, $I(\pi/2)$ can
change the sign form negative to positive with increasing
$\epsilon_{0}$. Then another $pi$-phase transition occurs as
shown in Fig. 2. When $\epsilon_{0}$ is comparable and close to
$r$, a maximum of $I(\pi/2)$ appears. The reason is related to
the position shift of the Andreev bound states with
$\epsilon_{0}$ as mentioned above. With increasing
$\epsilon_{0}$, $A_{1}$ and $A_{2}$ move to below the Fermi level
while $A_{-1}$ and $A_{-2}$ move to above the Fermi level. The
components contributed to the supercurrent change from ($A_{1}$,
$A_{-2}$, $C_{-1}$) to ($A_{1}$, $A_{2}$, $C_{-1}$), and then the
current first increases. With further increasing $\epsilon_{0}$,
the current density is suppressed greatly due to the asymmetry of
the intradot levels, and then the current decrease again.

\section{Conclusion}
In summary, by using the nonequilibrium Green's function method,
the spin-flip scattering effects on the supercurrent and Andreev
bound states are studied in detail. The supercurrent is mostly
contributed by the discrete Andreev bound states if there is no
spin-flip scattering. The original Andreev bound state is split
into two ones due to the spin-flip scattering, and the successive
two bound states carrying currents with opposite signs move to
the same direction with increasing spin-flip scattering strength.
The main contributions to supercurrents can be changed from the
discrete spectrum to the continuous spectrum at proper spin-flip
scattering strength, which results in the $\pi$-junction
transition. Furthermore, another $\pi$-junction transition can
appear if the intradot energy level is controlled to some proper
value by the gate voltage, because the successive two bound
states carrying currents with the same signs move to the same
direction. The main contributions to supercurrents are from the
discrete spectrum again, which results in this transition.
Although the strength of the spin-flip scattering and the
position of the intradot energy level have quite different
influence on the supercurrents, the two $\pi$-junction transition
mechanisms are involved in the change of the current density.

\section*{Acknowledgments}
This project is supported by NSFC under Grants No. 90103027 and
No. 50025206, and the National "973" Projects Foundation of China
(No. 2002CB613505).



\begin{center}
{\bf Figure Captions}
\end{center}

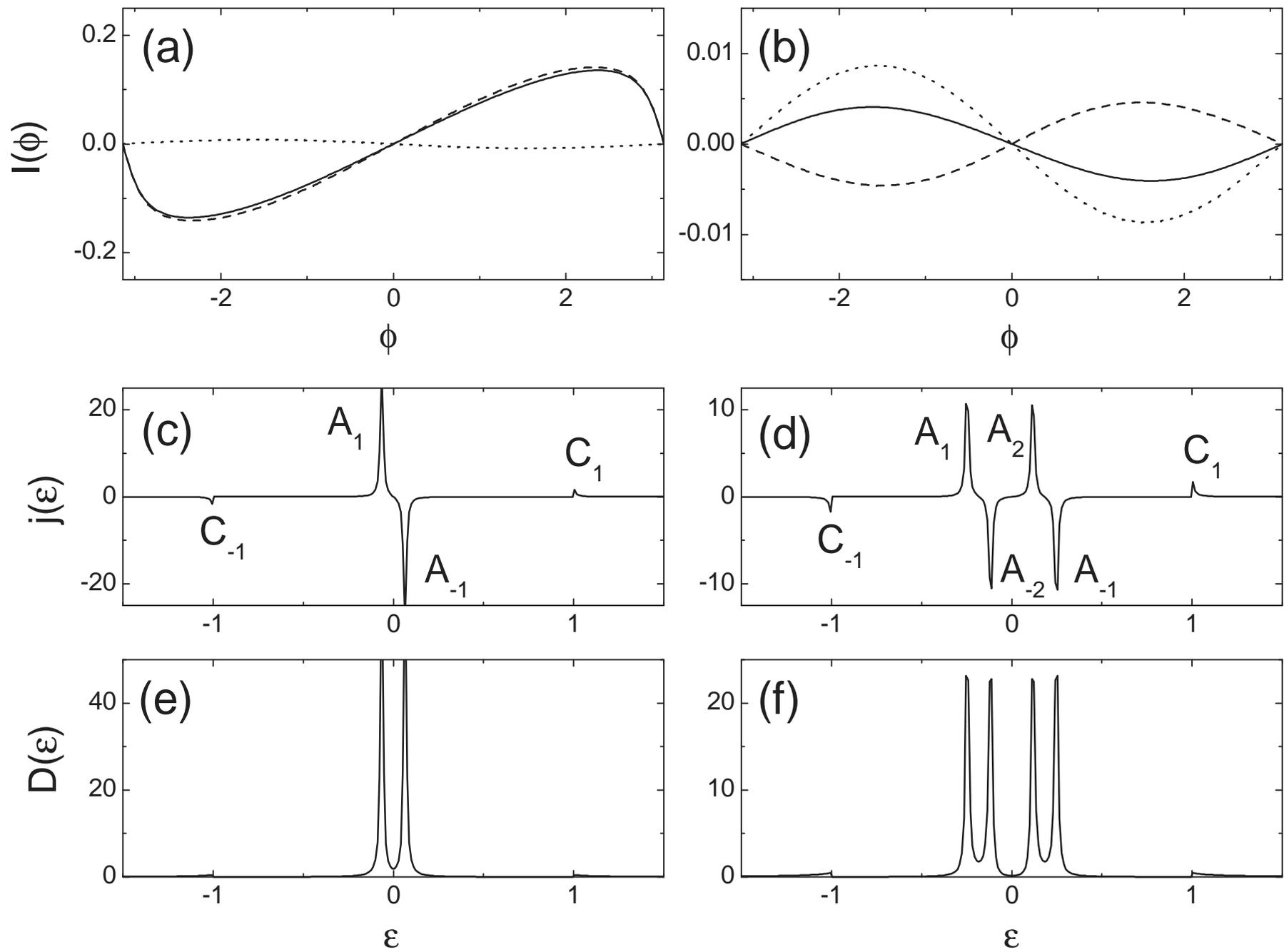
Fig. 1. The current $I$ (solid line), $I_{d}$ (dashed line), and
$I_{c}$ (dotted line) vs $\phi$ for $r=0$ (a) and $r=0.2$ (b) with
the dot level $\epsilon_{0}=0$. (c) and (d) are the corresponding
$j(\epsilon)$, and (e) and (f) are the corresponding
$D(\epsilon)$, respectively.

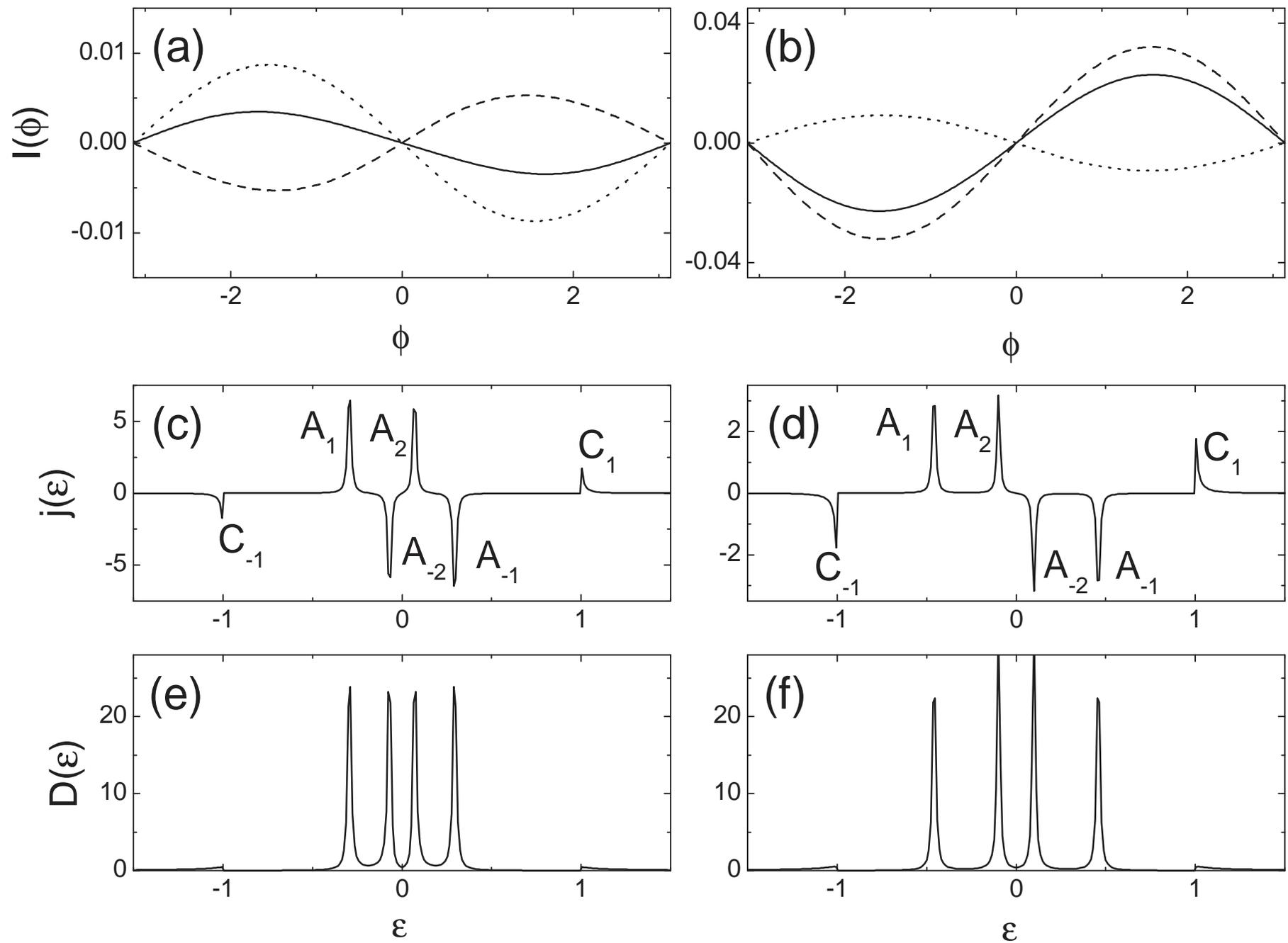
Fig. 2. The current $I$ (solid line), $I_{d}$ (dashed line), and
$I_{c}$ (dotted line) vs $\phi$ for $\epsilon_{0}=0.1$ (a) and
$\epsilon_{0}=0.2$ (b) with the spin-flip scattering strength
$r=0.2$. (c) and (d) are the corresponding $j(\epsilon)$, and (e)
and (f) are the corresponding $D(\epsilon)$, respectively.

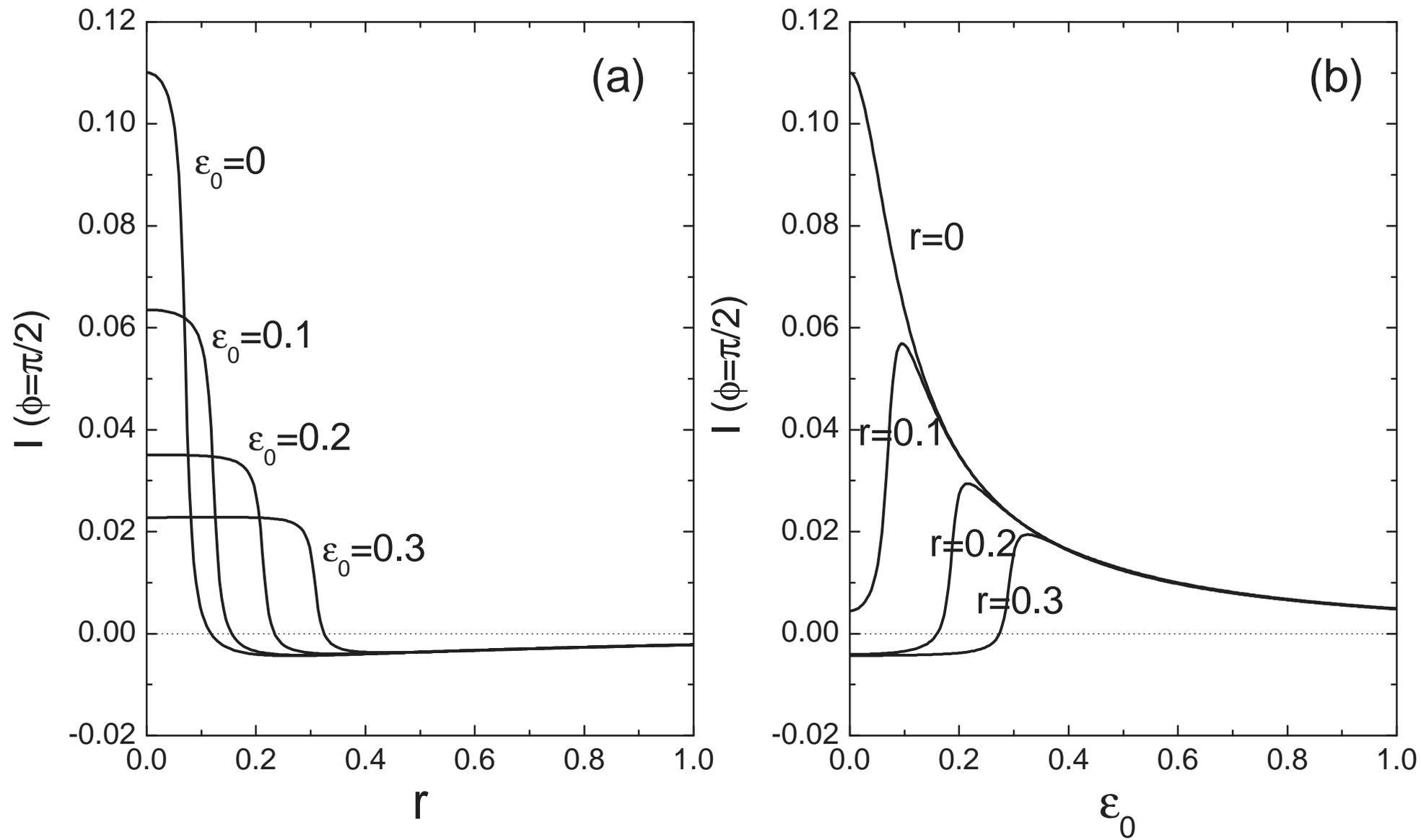
Fig. 3. (a) The current $I(\pi/2)$ vs $r$ at different
$\epsilon_{0}=0$, $0.1$, $0.2$ and $0.3$, respectively. (b) The
current $I(\pi/2)$ vs $\epsilon_{0}$ at different $r=0$, $0.1$,
$0.2$ and $0.3$, respectively.

\end{document}